\documentclass{article}

\usepackage{arxiv}

\usepackage[utf8]{inputenc} 
\usepackage[T1]{fontenc}    
\usepackage{hyperref}       
\usepackage{url}            
\usepackage{booktabs}       
\usepackage{amsfonts}       
\usepackage{nicefrac}       
\usepackage{microtype}      
\usepackage{lipsum}
\usepackage{graphicx}
\usepackage{algorithm}
\usepackage{algorithmic}
\usepackage{amsmath}
\usepackage{subcaption}
\usepackage{tabularx}
\usepackage[table]{xcolor}
\usepackage{caption}
\usepackage{bbm}
\usepackage{booktabs}
\usepackage{siunitx}
\usepackage{threeparttable}
\usepackage{siunitx,tabularx,booktabs,threeparttable}
\usepackage[authoryear]{natbib}
\usepackage[section]{placeins}
\usepackage{flafter}
\usepackage{dsfont}

\setkeys{Gin}{width=\linewidth, height=0.55\textheight, keepaspectratio}

\makeatletter
\def\fps@figure{tbp}
\makeatother

\usepackage{float}

\title{The Diversity Paradox revisited: Systemic Effects of Feedback Loops in Recommender Systems}

\date{}

\AtBeginDocument{%
  \addtocontents{toc}{}
  \pagestyle{plain}     
}

\author{
 Gabriele Barlacchi \\
  Scuola Normale Superiore and Università di Pisa\\
  Pisa, Italy \\
  \texttt{gabriele.barlacchi@sns.it} \\
   \And
 Margherita Lalli \\
  Scuola Normale Superiore\\
  Pisa, Italy \\
  \texttt{margherita.lalli@sns.it} \\
  \And
 Emanuele Ferragina \\
  Sciences Po\\
  Paris \\
  \texttt{emanuele.ferragina@sciencespo.fr} \\
  \And
 Fosca Giannotti \\
  Scuola Normale Superiore\\
  Pisa, Italy \\
  \texttt{fosca.giannotti@sns.it} \\
  \And
  Dino Pedreschi \\
  Università di Pisa\\
  Pisa, Italy \\
  \texttt{fosca.giannotti@sns.it} \\
  \And
 Luca Pappalardo \\
  CNR and Scuola Normale Superiore\\
  Pisa, Italy \\
  \texttt{luca.pappalardo@isti-cnr.it} \\
}

\begin{document}
\maketitle

\begin{abstract}
Recommender systems shape individual choices through feedback loops in which user behavior and algorithmic recommendations coevolve over time. 
The systemic effects of these loops remain poorly understood, in part due to unrealistic assumptions in existing simulation studies.
We propose a feedback-loop model that captures implicit feedback, periodic retraining, probabilistic adoption of recommendations, and heterogeneous recommender systems. 
We apply the framework on online retail and music streaming data and analyze systemic effects of the feedback loop.
We find that increasing recommender adoption may lead to a progressive diversification of individual consumption, while collective demand is redistributed in model- and domain-dependent ways, often amplifying popularity concentration. Temporal analyses further reveal that apparent increases in individual diversity observed in static evaluations are illusory: when adoption is fixed and time unfolds, individual diversity consistently decreases across all models.
Our results highlight the need to move beyond static evaluations and explicitly account for feedback-loop dynamics when designing recommender systems.
\end{abstract}


\section{Introduction}
\label{sec:introduction}
Recommender systems constitute a pervasive layer of algorithmic mediation that shapes how individuals make choices on online platforms.
For example, in the online retail domain, they suggest products that individuals are likely to purchase, influencing the propensity to buy. In music streaming, they curate personalized listening experiences, influencing how individuals are exposed over time to  artists, genres, and catalogs.

Since recommender systems are powered by machine learning models trained on user interaction data, they automatically unleash a \emph{feedback loop}, which can be defined as follows: individuals’ choices determine the data used to train the recommender system, and the recommender system’s outputs, in turn, influence subsequent choices \cite{pedreschi2025human}.
This recursive process unfolds over time, generating a coevolution cycle between individuals and algorithms.

The empirical study of feedback loops is challenging due to limited access to proprietary data, short observation horizons, and poor reproducibility.
As a result, they are often investigated through simulation-based approaches, which enable mechanism isolation, long-term analysis, and counterfactual evaluation under controlled conditions \cite{FERRERO_EFFECTS, cai2025agenticfeedbackloopmodeling, mansoury2020feedback, chaney2018algorithmic, MAURO_FEEDBACKLOOP}.
However, despite the publication of a growing body of simulation studies, our understanding of the \emph{systemic effects} of feedback loops remains fragmented \cite{pappalardo2025surveyimpactsrecommendersystems}.

From a methodological perspective, many approaches rely on simplifying assumptions with limited realism. They include the exclusive use of explicit feedback, such as ratings, the absence of periodic retraining of the recommender system, and the consideration of only a narrow set of recommender algorithms \cite{pappalardo2025surveyimpactsrecommendersystems}.
These choices abstract away key features of real-world online platforms, where user feedback is predominantly implicit, models are frequently retrained, and algorithms are heterogeneous.

Beyond methodological limitations, a central conceptual tension concerns the so-called \emph{diversity paradox} \cite{pappalardo2025surveyimpactsrecommendersystems, yi2022recommendation}.
Some studies report a paradoxical pattern in which recommender systems increase the diversity of individual consumption while simultaneously reducing the diversity of collective consumption \cite{hosanagar_2009, MAURO_FEEDBACKLOOP, matt2019factual}.
Others instead highlight the emergence of filter bubbles that progressively decrease individual diversity \cite{pappalardo2025surveyimpactsrecommendersystems, anderson2020spotify}.
Whether these seemingly conflicting findings reflect genuine trade-offs, domain-specific effects, or artifacts of how diversity is measured and compared remains poorly understood in the literature. This is particularly true with respect to the role of algorithmic influence and temporal dynamics \cite{pappalardo2025surveyimpactsrecommendersystems, fairness_contrary, hosanagar_2009, HAZRATI}.

This article introduces a feedback-loop model and a flexible simulation framework designed to disentangle the static and dynamic effects of recommender influence and to reassess the diversity paradox.
Our model captures repeated interactions between users and recommender systems, in which choices are driven by intrinsic preferences and algorithmic recommendations, interactions generate data, and models are periodically retrained on these data.
The framework supports probabilistic adoption of recommendations, allowing us to control the strength of algorithmic influence, and accommodates diverse recommender systems within a unified experimental setting.

We apply our model on two real-world datasets representative of distinct consumption settings -- online retail and music streaming. 
To quantify the systemic effects of the feedback loop, we measure changes in individual and collective consumption diversity, consumption concentration, and user similarity both across adoption rates and along the temporal evolution of the feedback loop.

Our results show that the diversity paradox largely arises from static, cross-sectional evaluations.
When comparing end-of-simulation outcomes across adoption rates, stronger recommender influence may appear to increase individual consumption diversity, while collective diversity often declines in a model- and domain-dependent manner.
However, temporal analyses reveal that this apparent individual-level diversification is transient. When the adoption rate is fixed and the feedback loop unfolds, individual consumption diversity consistently erodes over time across all recommender systems and datasets.
The adoption rate primarily affects the baseline level of diversity, not its long-term trajectory. We further show that these dynamics are underpinned by distinct mechanisms, leading to varying degrees of collective concentration and homogenization.

Our study highlights the need to move beyond static evaluations of recommender systems and to explicitly account for the long-term dynamics induced by feedback loops. Only by adopting a temporal perspective can we correctly interpret diversity-related effects and inform the design of recommender systems that balance short-term utility with long-term diversity and societal impact.

The remainder of the paper is organized as follows. Section \ref{sec:related_work} reviews the literature on the effects of feedback loops. Section \ref{sec:feedback_loop} describes our feedback-loop model, while Section \ref{sec:exp_setup} details its experimental instantiation on real-world datasets. 
Section \ref{sec:results} presents the experimental results and Section \ref{sec:discussion} discuss them in the light of the diversity paradox. 
Finally, Section
\ref{sec:conclusion} concludes the paper.

\subsubsection*{\bf Open source}
The code to replicate our study is available at:\\ \url{https://github.com/gabarlacchi/amazon_feedback_loop.git}

\section{Related Work}
\label{sec:related_work}
This section reviews research on how recommender systems influence user behavior in various domains, from online retail to social media. 
We group existing works into two approaches \cite{pappalardo2025surveyimpactsrecommendersystems}: \emph{online experiments} analyse real-world user interactions under controlled conditions and offer ecological validity, but face constraints from platform policies, short observation windows, and limited scalability; \emph{simulations} model user–system dynamics to explore hypothetical effects and counterfactual scenarios, but rely on assumptions that may limit external validity.

\textbf{Online experiments.} 
Empirical findings on consumption diversity are often conflicting. 
On the one hand, recommender systems can expand individual exploration but simultaneously concentrate demand at the aggregate level, the so-called diversity paradox \cite{fleder2009blockbuster,lee2018cross}. 
On the other hand, recommendations may narrow user choices, steering individuals toward more homogeneous item sets \cite{wang2024field} and reduce listening diversity over time \cite{anderson2020spotify,holtz2020engagement}.
Research has also identified mechanisms through which recommendations influence decisions, including effects on item discovery, purchase behavior, and quality perception with broader implications for willingness to pay and market concentration \cite{goldfarb2011online,aral2012information,elberse2008longtail, hinz2021causal}. 
Interface design moderates these effects, as shown in movie recommendation area \cite{sun2024interactive}. 
News recommendation research highlights conditional effects: algorithmic exposure can amplify polarization \cite{levy2021polarization} or reinforce filter bubbles depending on user predispositions and design choices \cite{knudsen2023news,loecherbach2021diverse,haim2018filterbubble}. 
Even large-scale experiments reach opposite conclusions: some find no short-term filter-bubble effects on political attitudes \cite{guess2023filterbubble} while others document increased exposure to extreme content through recommendation chains \cite{ribeiro2020radicalization}.

\textbf{Simulations.} 
By isolating mechanisms and testing counterfactuals, simulations reveal feedback dynamics unfolding over extended periods.
Early work demonstrates that algorithmic reinforcement narrows exposure and reduces system-level diversity \cite{jiang2019degenerate} and repeated recommendations lead to behavioral homogenization. 
Mansoury et al. \cite{mansoury2020feedback} formalize how sustained exposure shapes user preferences rather than merely reflecting them. Agent-based models \cite{chaney2018algorithmic,mansoury2020popularity} reveal how small initial biases propagate into large-scale imbalances through mutual user-algorithm adaptation.

Beyond digital contexts, recent work shows that recommendation feedback loops extend into urban spatial behaviour \cite{zeng2021recwalk,sanchez,MAURO_FEEDBACKLOOP}. 
Sanchez et al. \cite{sanchez} and Mauro et al. \cite{MAURO_FEEDBACKLOOP} demonstrate that repeated exposure to location-based recommendations systematically reshapes individual mobility choices, leading to long-term spatial concentration and collective homogenization.

Simulations also serve as testbeds for interventions. Helberger et al. \cite{helberger2019diversity} propose frameworks to balance diversity and autonomy in news, and Coppolillo et al. \cite{ALGO_DRIFT} proposes a framework for systematically evaluating recommender systems, with a particular focus on measuring drift in behavioral patterns and biases.
Interface design studies demonstrate how presentation shapes exposure  \cite{2022carousel}. Diversity-aware ranking and nudging strategies have been tested in simulation before deployment \cite{hazrati2021bias,mattis2022bias}.

Bias amplification under feedback loops has received particular attention. 
Hazrati and Ricci \cite{hazrati2021bias}  show how user biases distort evaluation metrics, with Mattis et al. \cite{mattis2022bias} demonstrating worsening effects over time. To address these risks, Hansen et al. \cite{hansen2023feedback} introduce various debiasing techniques.

Finally, modeling choices critically shape simulation outcomes. 
Zeng et al. \cite{zeng2020behavioral} demonstrate that behavioral model selection (e.g., multinomial logit vs. bounded rationality) significantly alters diversity predictions; and Mungari et al. \cite{manco_kdd} introduce a flexible framework for generating synthetic preference data to support recommendation analysis, showing the differences among all the possible approaches.

Collectively, these contributions establish simulation as a bridge between theory and practice, enabling study of long-term dynamics hidden in short-term evaluations.

\section{Modelling the Feedback Loop}
\label{sec:feedback_loop}
We aim to model a scenario in which a recommender system suggests items to individuals and iteratively adapts to their choices, giving rise to a feedback loop.

This section formally describes the model and its implementation as a simulation framework.
The model specifies the core components of the feedback loop -- scoring functions, recommendation policies, and user choice mechanisms -- together with their temporal coupling through repeated interaction and retraining cycles.

\subsection{Notation and Model componenets} \label{sec:notation}

Let $\mathcal{U}$ be a set of users, $\mathcal{I}$ a set of items, and $\mathcal{T} = {1, 2, \ldots, T}$ a sequence of discrete, ordered time steps.

We define a \emph{model scoring function}
$R: \mathcal{U} \times \mathcal{I} \times \mathcal{T} \rightarrow \mathbb{R}$
that assigns a real-valued relevance score $R(u,i,t)$ to each item $i \in \mathcal{I}$ for a user $u \in \mathcal{U}$ at time $t \in \mathcal{T}$.
This score represents the recommender system’s estimate of how relevant item $i$ is to user $u$ at time $t$.

The recommender system produces recommendations by selecting the highest-scoring items.
We define a \emph{recommender function} $\rho$ that maps a user $u$ and a time $t$ to a ranked list of $k$ items $K_{u,t}$: 
$$ \rho(u, t) := K_{u,t} = \underset{i \in \mathcal{I}}{\mathrm{top\text{-}k}} \, R(u,i,t) $$
\noindent where $\mathrm{top\text{-}k}$ returns the $k$ items with the highest scores, ordered in decreasing $R(u,i,t)$.

Users may also make choices autonomously, independently of the recommender system’s suggestions. 
In this case, their choices are drawn from a user- and time-specific candidate set $\mathcal{C}_{u,t} \subseteq \mathcal{I}$, representing the subset of items that user $u$ is aware of at time $t$. 
We model such autonomous choices through a \emph{user scoring function} $\hat{R}$, defined only over admissible user–item–time triplets: $$\hat{R}: \{ (u,i,t) \in \mathcal{U} \times \mathcal{I} \times \mathcal{T} \mid i \in \mathcal{C}_{u,t}  \} \rightarrow \mathbb{R}$$
\noindent For each triplet $(u,i,t)$, the function $\hat{R}(u,i,t)$ assigns a real-valued score that reflects the user’s intrinsic preference for item $i$ at time $t$.

\subsection{Feedback Loop dynamics}
\label{sec:dynamics}

The feedback loop model is implemented as a simulation framework, whose structure is described below.
See Algorithm 1 in Appendix A for the model's pseudocode.

\subsubsection* {\bf Initialization}
A real-world dataset describing user-item interactions is used to train the model scoring function $R$ and the user scoring function $\hat{R}$.
The simulation is initialized at time $t_0 \geq 1$ and proceeds by generating user choices at each subsequent discrete time step $t \in [t_0, T]$.
At each simulation step $t$, a subset of users $\mathcal{U}_t \subseteq \mathcal{U}$ becomes \emph{active}. 
Each active user $u \in \mathcal{U}_{t}$ selects a basket of items of size $b_{u,t}$.
Items in the basket are selected sequentially according to the item selection procedure.

\subsubsection*{\bf Item selection}
When selecting an item, an active user $u$ can choose among two mechanisms according to the adoption rate $\eta$:
\begin{itemize}
\item \textbf{Algorithmic choice}: with probability $\eta$, $u$ selects an item from the ranked list $K_{u,t}$ provided by the recommender system. 
This item is drawn from $K_{u,t}$ with a probability proportional to the scores $\{ R(u,i,t) \}_{i \in K_{u,t}}$.

\item \textbf{Autonomous choice}: with probability $1-\eta$, $u$ selects an item $i$ from the candidate set $\mathcal{C}_{u,t}$, according to a probability proportional to the score $\hat{R}(u,i,t)$ computed from the user scoring function.
\end{itemize}

\subsubsection*{\bf Recommender system update}
We partition the simulation timeline $[t_0, T]$ into discrete, non-overlapping intervals termed \emph{epochs}. 
Each epoch $j=1, \dots, n$ comprises a sequence of simulation steps.
At the end of each epoch $j$, the model scoring function $R$, the user scoring function $\hat{R}$, and the candidate set $\mathcal{C}_{u,t}$ are updated, and the recommender system is re-trained, inducing updated ranked lists $K_{u,t}$ for subsequent simulation steps.

\subsubsection*{\bf Autonomous Choice Model} \label{sec:user_choice_model}
We model autonomous choices following Cesa-Bianchi et al. and subsequent extensions \cite{BOLTZMAN, general_FL_1, feedback_sim_ricci, chaney2018algorithmic}.
The probability that an active user $u$ selects an item $i$ from its candidate set $\mathcal{C}_{u,t}$ at time $t$ is given by a softmax function:
\begin{equation} \label{eq: choice prob}
P(i \mid u, t) = \frac{\hat{R}(u,i,t)}{\sum_{j \in \mathcal{C}_{u,t}} \hat{R}(u,j,t)} = \frac{e^{V_{u,i}(t)/\tau}}{\sum_{j \in \mathcal{C}_{u,t}} e^{V_{u,j}(t)/\tau}},
\end{equation}
where $V_{u,i}(t)$ is an utility function associated with $i$ for $u$ at time $t$ and $\tau$ controls the exploration–exploitation balance.
As $\tau \rightarrow 0$, choices become deterministic toward the highest-utility item, while for $\tau \rightarrow \infty$, they approach a uniform distribution.

Following the approach by Hazrati et al. \cite{feedback_sim_ricci}, the candidate set for each user is constructed at each time by combining items from three sources: $GPop$ rank items by global popularity (i.e., the number of interactions by all users in the current epoch); $IPop$ rank items by individual popularity (i.e, the number of interactions from users' own purchase history up to the current epoch); $Unknown$ rank randomly items with which the user has had no prior interaction.
The final candidate set combines these sources with fixed proportions -- 40\% from $GPop$, 40\% from $IPop$, and 20\% from $Unknown$ -- balancing globally and individually popular items with exploration of unseen content.

We estimate the utility of item $i$ for user $u$ as adapting the approach by Mansoury et al. \cite{general_FL_1}:
\begin{equation}
V_{u,i}(t) = c_u (t) + G_u(t) \cdot \log(1 + s_i(t)) + \lambda \cdot \frac{1}{1 + s_i(t)} + \xi_{u,i}(t)
\end{equation}
where $c_u$ denotes the average number of interactions of user $u$. The term $G_u \cdot \log(1 + s_i)$ modulates the influence of item popularity $s_i$ based on the user’s interaction diversity $G_u$, computed as the Gini coefficient over the user's interaction distribution.  
The term $\lambda \cdot \frac{1}{1 + s_i}$ introduces a controlled boost for rare items, with $\lambda$ regulating the strength of this rarity effect \cite{HAZRATI, RARITY_1}. Finally, $\xi_{u,i}$ is a stochastic noise term, sampled at each timestep from a standard normal distribution. 
All the involved quantities ($c_u$, $G_u$, $s_i$) are evaluated by considering all simulation steps prior to $t$.

The autonomous choice model serves as a structured null model: more realistic than purely popularity-driven baselines, yet less arbitrary than random choice.

\section{Experimental Setup}
\label{sec:exp_setup}

We instantiate our model on two real-world datasets representative of distinct consumption settings: online retail and music streaming.

In this section, we introduce these datasets and preprocessing choices employed (Section \ref{sec:datasets});
we describe the set of benchmark recommender systems considered (Section \ref{sec:recommenders}); we detail the simulation protocol used to couple user choices and recommender updates over time (Section \ref{sec:protocol}); and we define the metrics employed to quantify systemic effects (Section \ref{sec:metrics}).

\subsection{Datasets}
\label{sec:datasets}
The two real-world datasets we consider describe which items individuals have consumed, without relying on explicit ratings.
We deliberately exclude rating-based datasets, as they record one-off evaluations rather than repeated consumption events, making them less suited to modeling longitudinal feedback loops. More importantly, large-scale online platforms are better characterized by implicit feedback \cite{aggarwal2016recommender}. In many domains -- such as music streaming or online retail -- individuals do not always provide ratings, while their behavior is continuously recorded through consumption events. 
As a result, focusing on implicit feedback allows us to study recommender effects in a more general setting than explicit-feedback-based approaches.
Finally, to the best of our knowledge, these two datasets represent the only publicly available, large-scale, and longitudinal sources of implicit feedback that support longitudinal analysis, further motivating their selection.

{\bf Amazon e-commerce 1.0.} 
This dataset contains anonymized purchase transactions collected from Amazon.com, where interactions represent item purchases without associated explicit ratings \cite{AMAZON_10}. The data has been collected through an ethical crowdsourcing process in which 5027 U.S. users voluntarily shared their detailed purchase histories as well as personal information including age, gender, education level, smoking habits, diabetes status, race, and self-reported frequency of Amazon usage. 
Item metadata consist of anonymized item identifiers and category labels provided by Amazon. 
The dataset contains 939,082 items belonging to 1871 categories.
We consider data covering a period from 2018 to 2020. 

{\bf Last.fm 1K.} In this dataset, interactions correspond to listening events \cite{lastfm}. The data were collected through the Last.fm APIs and consist of listening history records for 992 users, covering 960{,}402 music tracks and 107{,}295 albums. 
Available user-side information includes age, gender, and country, while item metadata capture track-to-album and track-to-artist associations. 
For the simulation, we consider data from 2006 to 2008.

For both datasets, we retain only users who are active for at least three months in each year. This constraint ensures temporal continuity and mitigates cold-start issues during the simulation. 
All interactions are timestamped and discretized at a daily granularity to align with the simulation steps within each epoch. 

After filtering, the Amazon dataset comprises 1{,}021 users and 1{,}448 items, while the Last.fm dataset contains 366 users and 9{,}254 artists.
Thus, user-item interactions in the Last.fm dataset are much sparser than those in the Amazon dataset.

\subsection{Benchmark recommender systems}
\label{sec:recommenders}

We compare a diverse set of recommender systems, spanning classical collaborative filtering methods and hybrid methods that combine collaborative signals with feature-based interactions. 
All recommender systems are implemented using the RecBole library \cite{recbole2.0}.

\subsubsection*{\bf Collaborative filtering methods} 
They leverage patterns in historical user–item interactions.
We select a representative set of algorithms spanning classical neighborhood-based methods, latent factor models, and graph-based neural approaches. 
Specifically, the set of algorithms includes:
\begin{itemize}
\item \textbf{ItemKNN} \cite{itemKNN} recommends items similar to those previously interacted with by the user, based on item--item co-occurrence similarity.
\item \textbf{BPR} \cite{BPR} learns personalized rankings via matrix factorization trained with a Bayesian pairwise ranking objective, contrasting observed and unobserved interactions.
\item \textbf{NeuMF} \cite{neuMF} combines generalized matrix factorization with multi-layer perceptrons to model both linear and nonlinear user--item interaction patterns.
\item \textbf{NGCF} \cite{ngcf} propagates embedding over the user-item interaction graph to exploit high-order connectivity and nonlinear transformations.
\item \textbf{SpectralCF} \cite{spectralCF}  operates in the spectral domain of the user--item graph to capture global structural patterns.
\item \textbf{SGL} \cite{sgl} enhances representation robustness through self-supervised learning by contrasting multiple perturbed views of the interaction graph.
\end{itemize}

\subsubsection*{\bf Hybrid methods}
They integrate interaction data with side information by explicitly modeling feature interactions. This allows us to study feedback-loop dynamics when recommendations are driven not only by co-occurrence patterns, but also by user- and item-side information.
We consider the following approaches:
\begin{itemize}
\item \textbf{DeepFM} \cite{deepFM} combines factorization machines and deep neural networks to jointly capture low-order feature interactions and high-order nonlinear patterns.
\item \textbf{DCN-V2} \cite{DCNV2} models bounded-degree feature crosses alongside deep representations, improving generalization in sparse and high-dimensional feature spaces.
\end{itemize}
To implement content-based information within hybrid approaches, user–item interactions are enriched with structured user and item side information contained in the two datasets.
Following RecBole preprocessing conventions, categorical features are tokenized and mapped to learned embedding vectors; numerical features are normalized and treated as continuous inputs; and textual item attributes (e.g., product categories) are represented as token sequences.
 
Table~\ref{tab:rec-results-datasets} in the Appendix summarizes the predictive accuracy of the recommender systems on the Amazon and Last.fm datasets, computed using standard evaluation metrics, while Table~\ref{tab:best-hparams_amazon} and Table~\ref{tab:best-hparams_lastfm} report the corresponding hyperparameters. 
All accuracy results are obtained using the temporal holdout protocol described in Section~\ref{sec:protocol}.

The results highlight a domain dependence: on Amazon, graph-based and representation-learning models such as SGL, NeuMF, and NGCF achieve the highest accuracy, whereas on Last.fm classical collaborative filtering approaches like BPR and ItemKNN perform better.
We note that the hybrid methods (DeepFM and DCN-V2) do not outperform the other models, suggesting that incorporating both user- and item-side information does not yield increased predictive power.
No single model consistently dominates across datasets and metrics; rather, all models exhibit competitive performance, justifying their inclusion in our analyses.

\subsection{Simulation protocol}
\label{sec:protocol}

We conduct independent simulations for each combination of dataset and recommender system. For each configuration, the model scoring function $R$ is initialized using a \emph{temporal holdout} protocol on the data: interactions from the first four months are used for training, the fifth month for validation, and the sixth month for testing. Hyperparameters are selected via grid-search optimization on the validation split and then fixed for the entire simulation.

The simulation timeline is divided into monthly epochs, starting from the seventh month of the real-world dataset and spanning two years, with simulation steps within each epoch corresponding to days in the real dataset.
At the end of each epoch, the model scoring function $R$ is updated using all interactions generated up to that point. 
To limit computational cost, after the first simulation year retraining is performed using a sliding temporal window that retains interactions from the most recent twelve months, together with the fixed initialization data.

A key control parameter of the feedback loop is the \emph{adoption rate} $\eta$, which determines the probability that an active user $u$ follows the algorithmic choice mechanism during item selection. We consider $\eta \in \{0, 0.2, 0.5, 0.8, 1\}$, spanning scenarios from no adoption ($\eta=0$) to full reliance on recommendations ($\eta=1$). 

When algorithmic choice occurs, the user selects an item from the ranked list $K_{u,t}$ produced by the recommender system. 
We fix $|K_{u,t}| = 20$ for all users and time steps. 
Item scores $\{R(u,i,t)\}_{i \in K{u,t}}$ are normalized to define a probability distribution, from which a single item is sampled.
At each time step $t$, the number of items $b_{u, t}$ simulated for a user is set equal to the number of interactions that the same user performs in the real dataset at that day.

In the autonomous choice model, each user is assigned a value of $\tau$ representing their exploration rate observed in the training data, estimated as the slope of the curve describing the number of new items consumed by the user over time.

To account for stochastic variability arising from user activation and item selection, each configuration is repeated over five independent runs.

\subsection{Quantifyng systemic effects}
\label{sec:metrics}
We adopt complementary metrics to capture the multidimensional systemic effects of the feedback loop: individual and collective consumption diversity, consumption concentration, and user similarity.
This choice is motivated by a recent survey \cite{pappalardo2025surveyimpactsrecommendersystems} showing that these measures are widely used across domains, though often in isolation and with fragmented results.

\subsubsection*{\bf Individual diversity}
We represent a user's consumption behavior through a vector $\mathbf{w}_u = \{w_{u,i} \mid i \in \mathcal{I}_u\}$, where $w_{u,i}$ denotes the number of times user $u$ has consumed item $i$, and $\mathcal{I}_u$ is the set of items appearing in $u$’s consumption history.  
To quantify how unevenly a user’s consumption is distributed across items, we compute the Gini coefficient of $\mathbf{w}_u$:
\begin{equation}\label{eq:Gini}
    G_u = \frac{\sum_{i=1}^{d_u} \sum_{j=1}^{d_u} |w_{u, i} - w_{u, j}|}{2 d_u^2 \overline{\mathbf{w}}_u}
\end{equation}
\noindent where $d_u = |\mathcal{I}_u|$ is the number of distinct items $u$ has consumed, and $\overline{\mathbf{w}}_u$ is the average number of consumptions per item.
We summarize individual consumption diversity at the population level by the mean Gini coefficient $\overline{G}_{\mathrm{ind}}$ computed over all users.

\subsubsection*{\bf Collective diversity}
We define a vector $\mathbf{s}$ whose components represent the total consumption volume of each item:
\begin{equation}\label{eq:prod_strength}
\mathbf{s}  = \{ s_{i} \mid i \in \mathcal{I} \}, \quad \text{where} \; s_i = \sum_{u\in\mathcal{U}} w_{u,i}.
\end{equation}
Here, $s_i$ denotes the overall popularity of item $i$ in the system, i.e., how many times it has been consumed.
We then compute the collective Gini coefficient $G_{\mathrm{coll}}$ by applying Equation~\ref{eq:Gini} to the system-level distribution $\mathbf{s}$, replacing the individual consumption vector $\mathbf{w}_u$, the number of distinct items $d_u$, and the mean consumption count $\overline{\mathbf{w}}_u$ with their system-level counterparts: $\mathbf{s}$, the total number of items $|\mathcal{I}|$, and the mean consumption volume per item $\overline{\mathbf{s}}$.

\subsubsection*{\bf Concentration}

We examine rank–frequency distributions at both the item and user levels.
Specifically, we consider the item-level distribution, defined by the item consumption volume $s_i$, and the user-level distribution, defined by the item popularity $p_i$:
\begin{equation}
\label{eq:item_pop}
p_i = \sum_{u \in \mathcal{U}} \mathds{1}{{w_{u,i} > 0}} ,
\end{equation}
where $\mathds{1}_{\{w_{u,i} > 0\}} = 1$ if user $u$ consumed $i$ at least once, and $0$ otherwise.
The item-level distribution captures concentration in terms of total interaction volume, while the user-level distribution reflects how many distinct users contribute to that volume.

\subsubsection*{\bf User similarity}
We quantify the similarity between the interaction histories of two users $u, v \in \mathcal{U}$ using the Jaccard index:
\begin{equation}
J(u,v) = \frac{|\mathcal{I}_u \cap \mathcal{I}_v|}{|\mathcal{I}_u \cup \mathcal{I}_v|}
\end{equation}
which ranges from $0$ (users have no items in common), to $1$ (identical interaction sets).

To obtain a system-level measure of user similarity, we compute the average Jaccard index across all unordered pairs of distinct users:
\begin{equation}
\overline{J}(\mathcal{U}) = \frac{1}{|\mathcal{P}(\mathcal{U})|} \sum_{(u,v) \in \mathcal{P}(\mathcal{U})} J(u,v),
\end{equation}
where $\mathcal{P}(\mathcal{U})$ denotes the set of all unordered user pairs.

\section{Results}
\label{sec:results}
We analyze the results along two complementary dimensions: \emph{(i)} variation with the adoption rate, capturing the macroscopic impact of increasing reliance on recommendations; and \emph{(ii)} temporal evolution at fixed adoption rate, revealing how feedback-loop effects emerge and accumulate over time.

\subsection{Systemic effects of Adoption Rate}
\label{sec:adoption_rate_effects}

\begin{figure}[htb!]
    \centering
    \includegraphics[width=\linewidth]{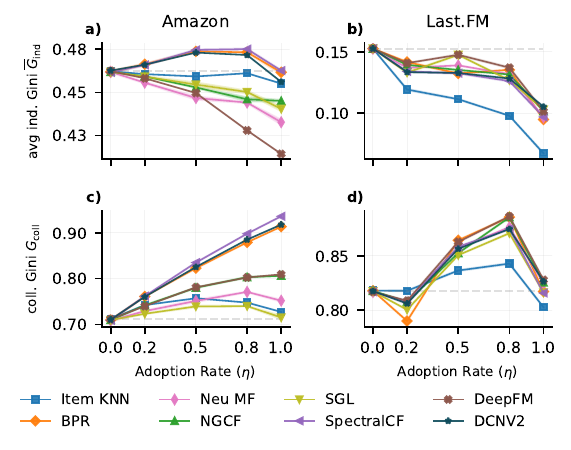}
    \caption{Variation of the average individual Gini (a, b) and the collective Gini (c, d) as a function of the adoption rate $\eta$, for the Amazon dataset (a, c) and the Last.fm dataset (b, d), across all benchmark recommender systems.}
    \label{fig1:indVScoll diversity}
\end{figure}

For different values of the adoption rate $\eta$, we compute all metrics at the end of the 24-epoch simulation. We then analyze how the metrics vary as a function of $\eta$.
\subsubsection*{\bf Individual diversity} 

Figure~\ref{fig1:indVScoll diversity}a-b show how the average individual Gini, $\overline{G}_{\mathrm{ind}}$, varies with the adoption rate $\eta$.
If the feedback loop had a negligible effect on individual consumption, $\overline{G}_{\mathrm{ind}}$ would remain constant as $\eta$ increases. 
Instead, for most recommender systems and across both datasets, $\overline{G}_{\mathrm{ind}}$ varies with $\eta$.

In the Amazon dataset, the variation in $\overline{G}_{\mathrm{ind}}$ is mild (see Figure \ref{fig1:indVScoll diversity}a).
Some models exhibit a weak inverted U-shaped trend, with a slight increase in $\overline{G}_{\mathrm{ind}}$ (i.e., a slight decrease in consumption diversity) at intermediate adoption rates that is reversed at high $\eta$.
Others instead display a clearer, approximately linear decrease in $\overline{G}_{\mathrm{ind}}$, indicating a more pronounced diversification effect.

In contrast, the Last.fm dataset shows a sharp decrease in $\overline{G}_{\mathrm{ind}}$ for all recommender systems (see Figure \ref{fig1:indVScoll diversity}b).
This means that individual consumption diversity at the end of the 24-epoch simulation increases as the adoption rate increases.
Differences across models affect the magnitude of the variation rather than its qualitative trend.
Notably, ItemKNN induces the strongest reduction in $\overline{G}_{\mathrm{ind}}$, highlighting a much stronger impact on individual diversity than observed in the Amazon dataset.
This indicates that the effect of the same recommender system can vary across domains.

\subsubsection*{\bf Collective diversity}
The collective Gini coefficient, $G_{\mathrm{coll}}$, exhibits markedly different behaviors across datasets (Figure \ref{fig1:indVScoll diversity}c-d).

In the Amazon dataset, $G_{\mathrm{coll}}$ generally increases with $\eta$, indicating that the feedback loop drives the system toward lower collective consumption diversity (see Figure \ref{fig1:indVScoll diversity}c).
This variation is substantial for several recommender systems; for example, SpectralCF rises from $G_{\mathrm{coll}}=0.71$ at $\eta=0$ to $0.95$ at $\eta=1$ (a 33\% increase).
While most recommender systems display a sublinear increase, SpectralCF, DCNV2, and BPR exhibit a sharp linear growth in $G_{\mathrm{coll}}$.
Notably, these three models also exhibit the smallest gains in individual consumption diversity, suggesting a link between limited individual diversification and stronger collective concentration effects.

In the Last.fm dataset, the variation in $G_{\mathrm{coll}}$ with $\eta$ is comparatively mild (see Figure \ref{fig1:indVScoll diversity}d). 
Although most recommender systems exhibit a non-monotonic pattern, with higher $G_{\mathrm{coll}}$ at intermediate adoption rates and lower values at both low and high $\eta$, the overall magnitude of the change remains limited.
ItemKNN stands out by exhibiting lower values of $G_{\mathrm{coll}}$ than the other models, corresponding to higher collective consumption diversity.
Taken together with the individual-level results, this indicates that in the Last.fm domain ItemKNN yields the strongest diversity gains at both the individual and collective levels under full adoption.

\subsubsection*{\bf Concentration}
To better interpret the aggregate signal captured by the collective Gini coefficient, we analyze two complementary distributions: \emph{(i)} the number of purchases per item as a function of item rank (\emph{item-level distribution}), which captures how total consumption volume is allocated across the items; and \emph{(ii)} the number of distinct users interacting with each item versus rank (\emph{user-level distribution}), which reflects how broadly user attention is spread across items.

\begin{figure}
    \centering
    \includegraphics[width=\linewidth]{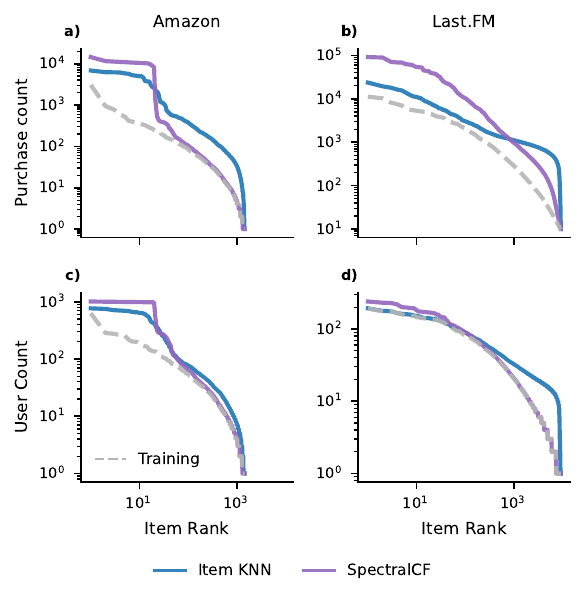}
\caption{(a,b) Item-level distribution, illustrating how total consumption volume is allocated across items. (c,d) User-level distribution, indicating how broadly user attention is spread across the catalog. Results for the Amazon and Last.fm datasets and for ItemKNN and SpectralCF.}
    \label{fig:Zipf}
\end{figure}

Figures~\ref{fig:Zipf}a–b show the item-level distributions for both datasets, comparing the training data with the full interaction histories obtained after the 24-epoch simulation under full adoption ($\eta=1$) for SpectralCF and ItemKNN.
These models are selected because they represent opposite extremes in the variation of $
G_{\mathrm{coll}}$ between $\eta=0$ and $\eta=1$.

In the Amazon dataset, recommendations markedly reshape the head of the distribution: top-ranked items accumulate substantially more purchases than in the training data, while the tail exhibits a faster decay (see Figure \ref{fig:Zipf}a). 
The two recommender systems differ in how they redistribute attention beyond the most popular items: ItemKNN increases interactions with mid-ranked and niche items, whose purchase volumes grow relative to the start of the simulation; SpectralCF concentrates demand almost exclusively on the top $k$ items, leaving medium- and niche-ranked items largely unchanged.

\begin{figure}[ht]
    \centering

    \begin{subfigure}[t]{0.48\linewidth}
        \centering
        \includegraphics[width=\linewidth]{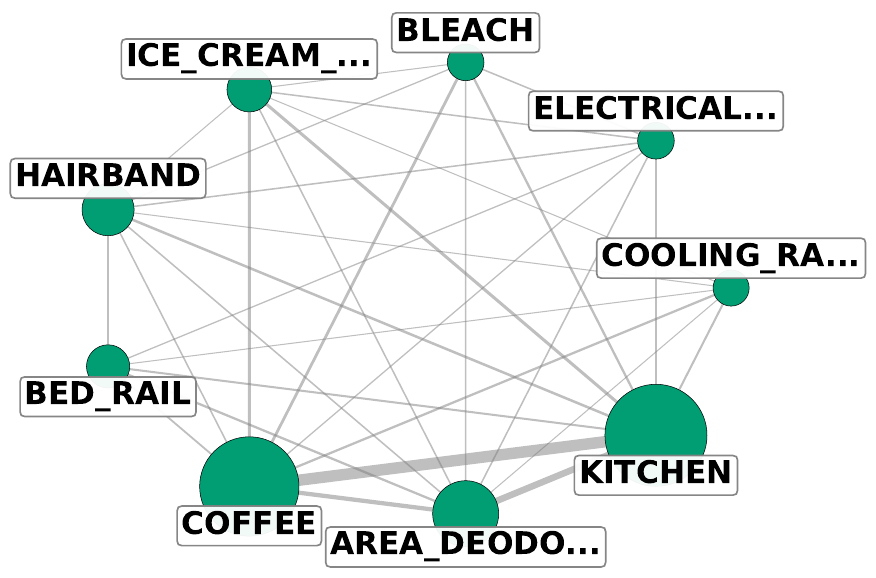}
        \caption{$\eta=0.2$}
        \label{fig:copurchase_network}
    \end{subfigure}
    \hfill
    \begin{subfigure}[t]{0.48\linewidth}
        \centering
        \includegraphics[width=\linewidth]{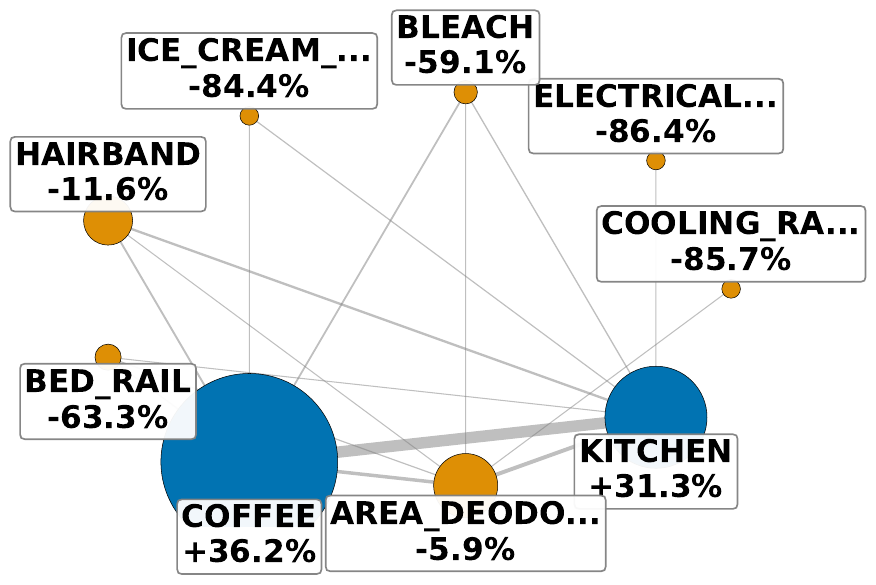}
        \caption{$\eta=0.8$}
        \label{fig1:copurchase_network_2}
    \end{subfigure}

\caption{Item co-purchase networks under $\eta{=}0.2$ and $\eta{=}0.8$ for SpectralCF.
Networks are extracted at the end of the simulation and focus on items with high, medium, and low popularity. In panel (b), edge colors indicate increases (green) or decreases (orange) in interaction volume, which is also showed through node size variation.}
    \label{fig:network_comparison}
\end{figure}

A concrete example of the concentration effect induced by SpectralCF is illustrated in Figure~\ref{fig:network_comparison}, which compares the co-purchase networks computed at the end of the 24-epoch simulation for $\eta=0.2$ and $\eta=0.8$. 
In these co-purchase networks, nodes represent items and edges denote co-purchase relationships.
Items are grouped by popularity into high-, medium-, and low-popularity classes, and the figure reports representative items from each group.
As $\eta$ increases, the co-purchase network becomes sparser and popular items become increasingly central, while less popular items lose visibility.
The percentage variation in interaction volume is encoded through changes in node size and color.
In this example, the most popular item at $\eta=0.8$ (“coffee”, green circle) gains 36\% more purchases than at $\eta=0.2$, while all other items lose up to 86.4\% (electrical outlet) of their purchases.

The Last.fm dataset shows a comparatively milder restructuring of the item-level distribution: although the head remains dominant, the discrepancy between the shape of the final and training distributions is smaller and the decay with rank is more gradual (Figure \ref{fig:Zipf}b). Notably, ItemKNN promotes lower-ranked items, producing a visible amplification of the tail.
This redirection of attention toward niche items is consistent with the low collective Gini coefficient observed for ItemKNN on Last.fm relative to other models.

The user-level distributions shown in Figure~\ref{fig:Zipf}c-d further clarify the concentration effect. In the Amazon dataset, recommendations increase the number of distinct users engaging with the top-ranked items, while mid- and low-ranked items attract fewer users or none at all (see Figure \ref{fig:Zipf}c). 
Thus, when users expand their consumption, they predominantly converge on already popular items. 
This effect is less pronounced for itemKNN, which also directs a small fraction of new interactions toward niche items, consistent with its modest changes in individual and collective Gini coefficients.

In contrast, the Last.fm user-level distributions closely track the training baseline across most ranks, indicating that the set of listeners associated with the high- and mid-popular songs remains essentially unchanged (see Figure \ref{fig:Zipf}d). 
The top-ranked tracks acquire virtually no new listeners and their observed increase in interaction volume is primarily driven by repeated engagement from existing users.
Note that while under ItemKNN niche items attract new listeners, SpectralCF produces a final distribution that is nearly indistinguishable from the initial one. 
Accordingly, the reduction in the individual Gini coefficient observed for SpectralCF should be interpreted not as evidence of genuine exploration, but rather as a reallocation of attention within users' item portfolios.


\subsubsection*{\bf User similarity}

\begin{figure}[htb!]
\centering   \includegraphics[width=\linewidth]{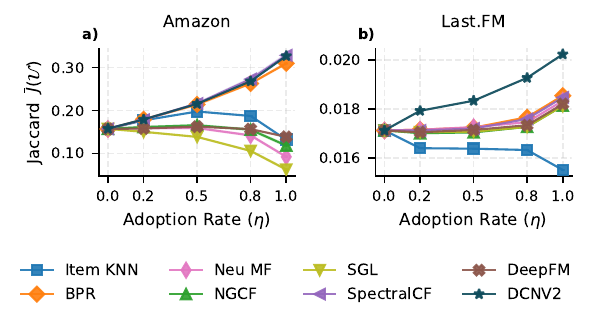}
\caption{Average user similarity $\overline{J}(\mathcal{U})$ as a function of the adoption rate $\eta$, for different recommender systems, on the Amazon (a) and Last.fm (b) datasets.}
\label{fig1:jaccard}
\end{figure}

Figure~\ref{fig1:jaccard}a–b reports the average user similarity $\overline{J}(\mathcal{U})$ as a function of $\eta$.
Overall, $\overline{J}(\mathcal{U})$ is substantially higher in the Amazon dataset than in Last.fm dataset. This reflects the greater sparsity of Last.fm, where users’ purchase histories overlap less due to a larger and more fragmented item space.

In the Amazon dataset, SpectralCF, BPR, and DCNV2 exhibit the strongest homogenizing effects: as $\eta$ approaches 1, $\overline{J}(\mathcal{U})$ increases superlinearly, more than doubling between no adoption ($\eta=0$) and full adoption ($\eta=1$).
This behavior mirrors the strong increase in the collective Gini coefficient observed for the same models, indicating that recommender systems which concentrate demand also induce homogenization of user behavior.

In the Last.fm dataset, $\overline{J}(\mathcal{U})$ varies within a much narrower range across adoption rates and recommender systems, with a maximum increase of about 15\% between $\eta=0$ and $\eta=1$.
ItemKNN induces almost no increase in similarity -- indeed, a slight decrease -- while DCNV2 shows the largest, yet still moderate, growth, consistent with the Amazon results.

Overall, these results show that user homogenization emerges from the interplay between recommender design, dataset structure, and feedback-loop dynamics, with some models inducing rapid convergence and others producing weak or saturating effects.

\subsection{Temporal dynamics of the Feedback Loop}
\label{sec:temporal_dynamics}

\begin{figure}[htb!]
    \centering
    \includegraphics[width=\linewidth]{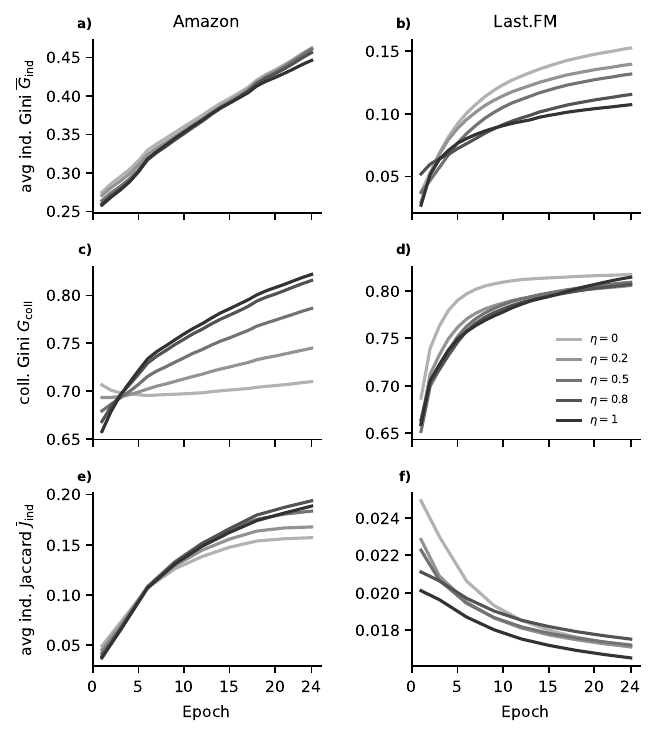}
    \caption{Evolution of the average individual Gini (a-b), collective Gini (c-d), and average individual similarity (e–f) over the 24 simulation epochs, for the Amazon and Last.fm datasets. Each curve corresponds to a different adoption rate $\eta$, averaged across recommender systems.}
        \label{fig:ginis_temporal}
\end{figure}

To investigate how systemic effects evolve as the feedback loop unfolds over time, we analyze the temporal dynamics of the metrics at fixed adoption rates. 
This perspective isolates the cumulative effects of recommender influence, allowing us to observe feedback-loop dynamics within each simulation run.

Figure \ref{fig:ginis_temporal} reports the evolution of the individual and collective Gini coefficients across the 24 simulation epochs for different values of $\eta$.
Each curve represents the average temporal trajectory across recommender systems. We adopt this aggregated view because the dynamics induced by different models are similar (see Figure \ref{fig1:ind_gini_epochs} in the Appendix) and averaging allows us to highlight the dominant temporal patterns without loss of information.

Across both datasets, $\overline{G}_{\mathrm{ind}}$ increases monotonically over epochs for all adoption rates, indicating a systematic reduction in individual consumption diversity as the feedback loop unfolds over epochs (Figure \ref{fig:ginis_temporal}a–b).
The adoption rate $\eta$ primarily affects the \emph{initial level} of the Gini coefficient: higher values of $\eta$ correspond to consistently lower $\overline{G}_{\mathrm{ind}}$ at the beginning of the simulation.
However, once the feedback loop starts, $\overline{G}_{\mathrm{ind}}$ follows a similar increasing trajectory over time regardless of $\eta$, showing that individual consumption diversity decreases monotonically even when stronger recommender influence initially induces more diversified consumption.

In the Amazon dataset, the curves corresponding to different adoption rates exhibit similar temporal shapes and follow an almost linear growth (see Figure \ref{fig:ginis_temporal}a).
In the Last.fm dataset, the temporal trajectories show slightly more variability across adoption rates and a clearly sublinear trend, with faster initial growth followed by gradual saturation (see Figure \ref{fig:ginis_temporal}b).
These differences are consistent with the static analysis of Section \ref{sec:adoption_rate_effects}: while Amazon shows limited sensitivity of individual consumption diversity to $\eta$, Last.fm exhibits stronger variation that nonetheless does not prevent a long-term erosion of individual diversity.

The temporal dynamics of the collective Gini coefficient reveal a nuanced picture.
In the Amazon dataset, different adoption rates induce qualitatively different temporal trajectories, with curves having different slopes (see Figure \ref{fig:ginis_temporal}c). Higher adoption rates lead to faster initial increases in concentration, followed by a slowdown.
In contrast, in the Last.fm dataset, $G_{\mathrm{coll}}$ evolves more consistently across adoption rates, with similar temporal shapes and smaller differences in magnitude (see Figure \ref{fig:ginis_temporal}d).
However, in the Last.fm dataset the role of the adoption rate in determining the initial positioning of the curve is less clear, consistent with the more irregular patterns observed in the dependence of $G_{\mathrm{coll}}$ on $\eta$.
In both datasets, however, the temporal trajectories of $G_{\mathrm{coll}}$ suggest diminishing returns: as the feedback loop unfolds, increases in concentration progressively slow down, indicating that the system approaches a saturation regime rather than diverging indefinitely.

Figure \ref{fig:ginis_temporal}e-f shows the evolution of user similarity over 24 simulation epochs for different values of $\eta$.
In the Amazon dataset, $\overline{J}(\mathcal{U})$ increases markedly over time, following a pattern of diminishing return. Higher adoption rates lead to stronger homogenization effects, with higher $\eta$ consistently producing higher similarity levels. 
The feedback loop increases user homogenization for all recommender systems, regardless of the adoption rate.
In contrast, in the Last.fm dataset the temporal variation in $\overline{J}(\mathcal{U})$ is mild, with an almost negligible decrease over time that also stabilizes. 
Here, higher adoption rates correspond to lower similarity curves, indicating slightly weaker homogenization effects, likely due to the much higher sparsity of the Last.fm dataset.

\section{Discussion}
\label{sec:discussion}
Our results provide a refined interpretation of the so-called \emph{diversity paradox}, often described as the coexistence of increasing individual consumption diversity and decreasing collective diversity under recommender influence \cite{pedreschi2025human, pappalardo2025surveyimpactsrecommendersystems, hosanagar_2009, MAURO_FEEDBACKLOOP, matt2019factual}. 
We show that this paradox largely arises from a static, cross-sectional comparison across adoption rates, rather than from the intrinsic dynamics of the feedback loop. 
When evaluated at the end of the simulation, higher adoption rates may appear to increase individual diversity while simultaneously reducing collective diversity.

A temporal perspective reveals a different mechanism. When the adoption rate is fixed and the feedback loop unfolds over time, individual diversity consistently decreases across all recommender systems and in both domains. 
\emph{The adoption rate mainly affects the initial position of the individual Gini curve, but not its temporal trend.} Although stronger recommender influence may initially induce more diversified consumption, the feedback loop progressively narrows individuals’ consumption set regardless of $\eta$.

Collective diversity decreases both when comparing outcomes across adoption rates and when tracking the system over time at fixed $\eta$. 
The underlying mechanisms, however, depend on recommender design and domain structure. In Amazon, concentration is driven by strong reinforcement of already popular items and increasing user homogenization. In Last.fm, changes are milder and often reflect repeated engagement within existing user–item portfolios rather than convergence toward the same items.

User similarity further supports this interpretation. In the Amazon dataset, similarity increases sharply over time, indicating that feedback loops align users’ consumption histories in dense interaction settings. In contrast, similarity varies only weakly in Last.fm, consistent with sparser data and weaker homogenization effects.

\section{Conclusion}
\label{sec:conclusion}
In this work, we studied the systemic effects of feedback loops in recommender systems and provided a new interpretation of the diversity paradox. 
Our study highlights the necessity to combine end-of-horizon and temporal evaluations frameworks to correctly assess the impact of recommender systems. 

A natural next step is to move from diagnosis to intervention. Future work should investigate how recommender systems can be designed to mitigate the long-term loss of diversity at both the individual and collective levels, while also limiting excessive concentration and user similarity induced by feedback loops. Addressing these challenges is essential for developing recommender systems that sustain exploration, preserve socio-diversity, and balance individual engagement with long-term systemic outcomes.

\bibliographystyle{plainnat}
\bibliography{reference}

\begin{thebibliography}{56}
\providecommand{\natexlab}[1]{#1}
\providecommand{\url}[1]{\texttt{#1}}
\expandafter\ifx\csname urlstyle\endcsname\relax
  \providecommand{\doi}[1]{doi: #1}\else
  \providecommand{\doi}{doi: \begingroup \urlstyle{rm}\Url}\fi

\bibitem[A.~Berke and Mahari(2024)]{AMAZON_10}
D.~Calacci A.~Berke and R.~Mahari.
\newblock Open e-commerce 1.0, five years of crowdsourced u.s. amazon purchase histories with user demographics, 2024.
\newblock URL \url{https://doi.org/10.1038/s41597-024-03329-6}.

\bibitem[Aggarwal(2016)]{aggarwal2016recommender}
Charu~C. Aggarwal.
\newblock \emph{Recommender Systems: The Textbook}.
\newblock Springer Publishing Company, Incorporated, 1st edition, 2016.
\newblock ISBN 3319296574.

\bibitem[Anderson et~al.(2020)Anderson, Maystre, Anderson, Mehrotra, and Lalmas]{anderson2020spotify}
Ashton Anderson, Lucas Maystre, Ian Anderson, Rishabh Mehrotra, and Mounia Lalmas.
\newblock Algorithmic effects on the diversity of consumption on spotify.
\newblock In \emph{Proceedings of The Web Conference 2020 (WWW '20)}, pages 2155--2165, USA, 2020. ACM.

\bibitem[Aral et~al.(2012)Aral, Brynjolfsson, and Alstyne]{aral2012information}
Sinan Aral, Erik Brynjolfsson, and Marshall~Van Alstyne.
\newblock Information, technology, and information worker productivity.
\newblock \emph{Information Systems Research}, 23\penalty0 (3):\penalty0 849--867, 2012.
\newblock \doi{10.2307/23274649}.

\bibitem[Bertin-Mahieux et~al.(2011)Bertin-Mahieux, Ellis, Whitman, and Lamere]{lastfm}
Thierry Bertin-Mahieux, Daniel~P.W. Ellis, Brian Whitman, and Paul Lamere.
\newblock The million song dataset.
\newblock In \emph{{Proceedings of the 12th International Conference on Music Information Retrieval ({ISMIR} 2011)}}, USA, 2011.

\bibitem[Cai et~al.(2025)Cai, Zhang, Bao, Gao, Wang, Feng, and He]{cai2025agenticfeedbackloopmodeling}
Shihao Cai, Jizhi Zhang, Keqin Bao, Chongming Gao, Qifan Wang, Fuli Feng, and Xiangnan He.
\newblock Agentic feedback loop modeling improves recommendation and user simulation, 2025.
\newblock URL \url{https://arxiv.org/abs/2410.20027}.

\bibitem[Cesa-Bianchi et~al.(2017)Cesa-Bianchi, Gentile, Lugosi, and Neu]{BOLTZMAN}
Nicolò Cesa-Bianchi, Claudio Gentile, Gábor Lugosi, and Gergely Neu.
\newblock Boltzmann exploration done right, 2017.
\newblock URL \url{https://arxiv.org/abs/1705.10257}.

\bibitem[Chaney et~al.(2018)Chaney, Stewart, and Engelhardt]{chaney2018algorithmic}
Allison J.~B. Chaney, Brandon~M. Stewart, and Barbara~E. Engelhardt.
\newblock How algorithmic confounding in recommendation systems increases homogeneity and decreases utility.
\newblock In \emph{Proceedings of the 12th ACM Conference on Recommender Systems (RecSys ’18)}, pages 224--232, 2018.
\newblock URL \url{https://dl.acm.org/doi/10.1145/3240323.3240370}.

\bibitem[Coppolillo et~al.(2024)Coppolillo, Mungari, Ritacco, Fabbri, Minici, Bonchi, and Manco]{ALGO_DRIFT}
Erica Coppolillo, Simone Mungari, Ettore Ritacco, Francesco Fabbri, Marco Minici, Francesco Bonchi, and Giuseppe Manco.
\newblock Algorithmic drift: A simulation framework to study the effects of recommender systems on user preferences, 2024.
\newblock URL \url{https://arxiv.org/abs/2409.16478}.

\bibitem[Deshpande and Karypis()]{itemKNN}
Mukund Deshpande and George Karypis.
\newblock Item-based top-n recommendation algorithms.
\newblock \emph{ACM Trans. Inf. Syst.}, 22\penalty0 (1):\penalty0 143–177.
\newblock URL \url{https://doi.org/10.1145/963770.963776}.

\bibitem[Elberse(2008)]{elberse2008longtail}
Anita Elberse.
\newblock Should you invest in the long tail?
\newblock \emph{Harvard Business Review}, 86\penalty0 (7/8):\penalty0 88--96, 2008.
\newblock URL \url{https://www.hbs.edu/faculty/Pages/item.aspx?num=32337}.

\bibitem[Fleder and Hosanagar(2009)]{fleder2009blockbuster}
Daniel Fleder and Kartik Hosanagar.
\newblock Blockbuster culture's next rise or fall: The impact of recommender systems on sales diversity.
\newblock \emph{Management Science}, 55\penalty0 (5):\penalty0 697--712, 2009.
\newblock \doi{10.1287/mnsc.1080.0974}.

\bibitem[Fleder et~al.(2010)Fleder, Hosanagar, and Buja]{hosanagar_2009}
Daniel Fleder, Kartik Hosanagar, and Andreas Buja.
\newblock Recommender systems and their effects on consumers: the fragmentation debate.
\newblock EC '10, page 229–230, New York, NY, USA, 2010. Association for Computing Machinery.
\newblock \doi{10.1145/1807342.1807378}.
\newblock URL \url{https://doi.org/10.1145/1807342.1807378}.

\bibitem[Goldfarb and Tucker(2011)]{goldfarb2011online}
Avi Goldfarb and Catherine Tucker.
\newblock Online display advertising: Targeting and obtrusiveness.
\newblock \emph{Marketing Science}, 30\penalty0 (3):\penalty0 389--404, 2011.
\newblock \doi{10.1287/mksc.1100.0583}.

\bibitem[Guess et~al.(2023)Guess, Nagler, and Tucker]{guess2023filterbubble}
Andrew Guess, Jonathan Nagler, and Joshua Tucker.
\newblock The filter bubble hypothesis: Experimental evidence.
\newblock \emph{Science Advances}, 9\penalty0 (14):\penalty0 eade0140, 2023.

\bibitem[Guo et~al.()Guo, Tang, Ye, Li, and He]{deepFM}
Huifeng Guo, Ruiming Tang, Yunming Ye, Zhenguo Li, and Xiuqiang He.
\newblock Deepfm: a factorization-machine based neural network for ctr prediction.
\newblock In \emph{Proceedings of the 26th International Joint Conference on Artificial Intelligence}, page 1725–1731. AAAI Press.

\bibitem[Haim et~al.(2018)Haim, Graefe, and Brosius]{haim2018filterbubble}
Mario Haim, Andreas Graefe, and Hans-Bernd Brosius.
\newblock Burst of the filter bubble? effects of personalization on the diversity of google news.
\newblock \emph{Digital Journalism}, 6\penalty0 (3):\penalty0 330--343, 2018.
\newblock \doi{10.1080/21670811.2017.1338145}.

\bibitem[Hansen et~al.(2023)Hansen, Mehrotra, McInerney, and Biega]{hansen2023feedback}
Casper Hansen, Rishabh Mehrotra, James McInerney, and Asia~J. Biega.
\newblock Correcting feedback loops in recommendation with simulation-based debiasing.
\newblock In \emph{Proceedings of the 17th ACM Conference on Recommender Systems (RecSys ’23)}, pages 411--421, 2023.

\bibitem[Hazrati and Ricci(2021)]{hazrati2021bias}
Maryam Hazrati and Francesco Ricci.
\newblock Bias amplification in recommender systems: Simulation study.
\newblock In \emph{Proceedings of the 29th ACM Conference on User Modeling, Adaptation and Personalization (UMAP ’21)}, pages 240--244, 2021.

\bibitem[Hazrati and Ricci()]{HAZRATI}
Naieme Hazrati and Francesco Ricci.
\newblock Recommender systems effect on the evolution of users’ choices distribution.
\newblock \emph{Information Processing and Management}, 59\penalty0 (1):\penalty0 102766.
\newblock URL \url{https://www.sciencedirect.com/science/article/pii/S0306457321002466}.

\bibitem[Hazrati et~al.()Hazrati, Elahi, and Ricci]{feedback_sim_ricci}
Naieme Hazrati, Mehdi Elahi, and Francesco Ricci.
\newblock Simulating the impact of recommender systems on the evolution of collective users' choices.
\newblock page 207–212, New York, NY, USA. Association for Computing Machinery.
\newblock URL \url{https://doi.org/10.1145/3372923.3404812}.

\bibitem[He et~al.()He, Liao, Zhang, Nie, Hu, and Chua]{neuMF}
Xiangnan He, Lizi Liao, Hanwang Zhang, Liqiang Nie, Xia Hu, and Tat-Seng Chua.
\newblock Neural collaborative filtering.
\newblock page 173–182, Republic and Canton of Geneva, CHE. International World Wide Web Conferences Steering Committee.
\newblock URL \url{https://doi.org/10.1145/3038912.3052569}.

\bibitem[Helberger et~al.(2019)Helberger, Karppinen, and D'Acunto]{helberger2019diversity}
Natali Helberger, Kari Karppinen, and Lucia D'Acunto.
\newblock On the democratic role of news recommenders.
\newblock \emph{Digital Journalism}, 7\penalty0 (8):\penalty0 993--1012, 2019.
\newblock URL \url{https://www.tandfonline.com/doi/full/10.1080/21670811.2019.1623700}.

\bibitem[Hinz et~al.(2021)Hinz, Li, and Grahl]{hinz2021causal}
Oliver Hinz, Xitong Li, and J{\"o}rn Grahl.
\newblock How do recommender systems lead to consumer purchases? a causal mediation analysis of a field experiment.
\newblock \emph{Information Systems Research}, 33\penalty0 (2):\penalty0 620--637, 2021.
\newblock \doi{10.1287/isre.2021.1074}.

\bibitem[Holtz et~al.(2020)Holtz, Carterette, Chandar, Nazari, Cramer, and Aral]{holtz2020engagement}
David Holtz, Ben Carterette, Praveen Chandar, Zahra Nazari, Henriette Cramer, and Sinan Aral.
\newblock The engagement-diversity connection: Evidence from a field experiment on spotify.
\newblock In \emph{Proceedings of the 21st ACM Conference on Economics and Computation (EC '20)}, pages 75--76. ACM, 2020.

\bibitem[Jiang et~al.(2019)Jiang, Chiappa, Lattimore, Gy{\"o}rgy, and Kohli]{jiang2019degenerate}
Ray Jiang, Silvia Chiappa, Tor Lattimore, Andr{\'a}s Gy{\"o}rgy, and Pushmeet Kohli.
\newblock Degenerate feedback loops in recommender systems.
\newblock In \emph{Proceedings of the AAAI Conference on Artificial Intelligence}, volume~33, pages 4275--4282, 2019.
\newblock URL \url{https://dl.acm.org/doi/10.1145/3306618.3314288}.

\bibitem[Knudsen(2023)]{knudsen2023news}
Erik Knudsen.
\newblock Modeling news recommender systems’ conditional effects on selective exposure: Evidence from two online experiments.
\newblock \emph{Journal of Communication}, 73\penalty0 (2):\penalty0 138--149, 2023.
\newblock URL \url{https://doi.org/10.1093/joc/jqac047}.

\bibitem[Lee and Hosanagar(2018)]{lee2018cross}
Dokyun Lee and Kartik Hosanagar.
\newblock How do recommender systems affect sales diversity? a cross-category investigation via randomized field experiment.
\newblock \emph{Information Systems Research}, 30\penalty0 (1):\penalty0 239--259, 2018.

\bibitem[Levy(2021)]{levy2021polarization}
Ro’ee Levy.
\newblock Social media, news consumption, and polarization: Evidence from a field experiment.
\newblock \emph{American Economic Review}, 111\penalty0 (3):\penalty0 831--870, 2021.

\bibitem[Loecherbach et~al.(2021)Loecherbach, Welbers, Moeller, Trilling, and Atteveldt]{loecherbach2021diverse}
Felicia Loecherbach, Kasper Welbers, Judith Moeller, Damian Trilling, and Wouter~Van Atteveldt.
\newblock Is this a click towards diversity? explaining when and why news users make diverse choices.
\newblock In \emph{Proceedings of the 13th ACM Web Science Conference 2021 (WebSci '21)}, pages 282--290. ACM, 2021.
\newblock URL \url{https://doi.org/10.1145/3447535.3462506}.

\bibitem[Mansoury et~al.(2020{\natexlab{a}})Mansoury, Abdollahpouri, Pechenizkiy, Mobasher, and Burke]{general_FL_1}
Masoud Mansoury, Himan Abdollahpouri, Mykola Pechenizkiy, Bamshad Mobasher, and Robin Burke.
\newblock Feedback loop and bias amplification in recommender systems, 2020{\natexlab{a}}.
\newblock URL \url{https://arxiv.org/abs/2007.13019}.

\bibitem[Mansoury et~al.(2020{\natexlab{b}})Mansoury, Abdollahpouri, Pechenizkiy, Mobasher, and Burke]{mansoury2020feedback}
Masoud Mansoury, Himan Abdollahpouri, Mykola Pechenizkiy, Bamshad Mobasher, and Robin Burke.
\newblock Feedback loops in recommender systems: Quantification and mitigation.
\newblock In \emph{Proceedings of the 29th ACM International Conference on Information and Knowledge Management (CIKM ’20)}, pages 2145--2148, 2020{\natexlab{b}}.
\newblock URL \url{https://dl.acm.org/doi/10.1145/3340531.3412152}.

\bibitem[Mansoury et~al.(2020{\natexlab{c}})Mansoury, Abdollahpouri, Pechenizkiy, Mobasher, and Burke]{mansoury2020popularity}
Masoud Mansoury, Himan Abdollahpouri, Mykola Pechenizkiy, Bamshad Mobasher, and Robin Burke.
\newblock Popularity bias in recommendation: A multi-stakeholder perspective.
\newblock In \emph{Proceedings of the 13th ACM Conference on Recommender Systems (RecSys ’20)}, pages 342--346, 2020{\natexlab{c}}.
\newblock URL \url{https://dl.acm.org/doi/10.1145/3383313.3412232}.

\bibitem[Matt et~al.(2019)Matt, Hess, and Wei{\ss}]{matt2019factual}
Christian Matt, Thomas Hess, and Christian Wei{\ss}.
\newblock A factual and perceptional framework for assessing diversity effects of online recommender systems.
\newblock \emph{Internet research}, 29\penalty0 (6):\penalty0 1526--1550, 2019.

\bibitem[Mattis et~al.(2022)Mattis, Burke, and Abdollahpouri]{mattis2022bias}
John Mattis, Robin Burke, and Himan Abdollahpouri.
\newblock Modeling user bias evolution in recommender systems.
\newblock In \emph{Proceedings of the 16th ACM Conference on Recommender Systems (RecSys ’22)}, pages 648--652, 2022.

\bibitem[Mauro et~al.(2026)Mauro, Minici, and Pappalardo]{MAURO_FEEDBACKLOOP}
Giovanni Mauro, Marco Minici, and Luca Pappalardo.
\newblock The urban impact of ai: modelling feedback loops in location-based recommender systems.
\newblock \emph{Mach. Learn.}, 115\penalty0 (1), January 2026.
\newblock ISSN 0885-6125.
\newblock \doi{10.1007/s10994-025-06904-z}.
\newblock URL \url{https://doi.org/10.1007/s10994-025-06904-z}.

\bibitem[Mungari et~al.(2025)Mungari, Coppolillo, Ritacco, and Manco]{manco_kdd}
Simone Mungari, Erica Coppolillo, Ettore Ritacco, and Giuseppe Manco.
\newblock Flexible generation of preference data for recommendation analysis.
\newblock page 5710–5721, New York, NY, USA, 2025. Association for Computing Machinery.
\newblock URL \url{https://doi.org/10.1145/3711896.3737398}.

\bibitem[Pappalardo et~al.(2024)Pappalardo, Ferragina, Citraro, Cornacchia, Nanni, Rossetti, Gezici, Giannotti, Lalli, Gambetta, Mauro, Morini, Pansanella, and Pedreschi]{pappalardo2025surveyimpactsrecommendersystems}
Luca Pappalardo, Emanuele Ferragina, Salvatore Citraro, Giuliano Cornacchia, Mirco Nanni, Giulio Rossetti, Gizem Gezici, Fosca Giannotti, Margherita Lalli, Daniel~Rodriguez Gambetta, Giovanni Mauro, Virginia Morini, Valentina Pansanella, and Dino Pedreschi.
\newblock A survey on the impacts of recommender systems on users, items, and human-ai ecosystems.
\newblock 2024.
\newblock URL \url{https://api.semanticscholar.org/CorpusID:270878532}.

\bibitem[Pedreschi et~al.(2025)Pedreschi, Pappalardo, Ferragina, Baeza-Yates, Barab{\'a}si, Dignum, Dignum, Eliassi-Rad, Giannotti, Kert{\'e}sz, et~al.]{pedreschi2025human}
Dino Pedreschi, Luca Pappalardo, Emanuele Ferragina, Ricardo Baeza-Yates, Albert-L{\'a}szl{\'o} Barab{\'a}si, Frank Dignum, Virginia Dignum, Tina Eliassi-Rad, Fosca Giannotti, J{\'a}nos Kert{\'e}sz, et~al.
\newblock Human-ai coevolution.
\newblock \emph{Artificial Intelligence}, 339:\penalty0 104244, 2025.

\bibitem[Rahdari et~al.(2024)Rahdari, Brusilovsky, and Kveton]{2022carousel}
Behnam Rahdari, Peter Brusilovsky, and Branislav Kveton.
\newblock Towards simulation-based evaluation of recommender systems with carousel interfaces.
\newblock \emph{ACM Transactions on Recommender Systems}, 2024.
\newblock \doi{https://doi.org/10.1145/3643709}.
\newblock To appear / published online (depends on version).

\bibitem[Rendle et~al.()Rendle, Freudenthaler, Gantner, and Schmidt-Thieme]{BPR}
Steffen Rendle, Christoph Freudenthaler, Zeno Gantner, and Lars Schmidt-Thieme.
\newblock Bpr: Bayesian personalized ranking from implicit feedback.
\newblock page 452–461, Arlington, Virginia, USA. AUAI Press.

\bibitem[Ribeiro et~al.(2020)Ribeiro, Ottoni, West, Almeida, and Jr.]{ribeiro2020radicalization}
Manoel~Horta Ribeiro, Raphael Ottoni, Robert West, Virg{\'i}lio A.~F. Almeida, and Wagner~Meira Jr.
\newblock Auditing radicalization pathways on youtube.
\newblock In \emph{Proceedings of the 2020 ACM Conference on Fairness, Accountability, and Transparency (FAT* '20)}, pages 131--141. ACM, 2020.
\newblock URL \url{https://doi.org/10.1145/3351095.3372879}.

\bibitem[S\'{a}nchez et~al.(2023)S\'{a}nchez, Bellog\'{\i}n, and Boratto]{sanchez}
Pablo S\'{a}nchez, Alejandro Bellog\'{\i}n, and Ludovico Boratto.
\newblock Bias characterization, assessment, and mitigation in location-based recommender systems.
\newblock \emph{Data Min. Knowl. Discov.}, 37\penalty0 (5):\penalty0 1885–1929, 2023.
\newblock \doi{10.1007/s10618-022-00913-5}.
\newblock URL \url{https://doi.org/10.1007/s10618-022-00913-5}.

\bibitem[Sonboli et~al.(2022)Sonboli, Burke, Ekstrand, and Mehrotra]{fairness_contrary}
Nasim Sonboli, Robin Burke, Michael Ekstrand, and Rishabh Mehrotra.
\newblock The multisided complexity of fairness in recommender systems.
\newblock \emph{AI Mag.}, 43\penalty0 (2):\penalty0 164–176, June 2022.
\newblock \doi{10.1002/aaai.12054}.
\newblock URL \url{https://doi.org/10.1002/aaai.12054}.

\bibitem[Sun et~al.(2024)Sun, Akella, Kong, Zhou, and Konstan]{sun2024interactive}
Ruixuan Sun, Avinash Akella, Ruoyan Kong, Moyan Zhou, and Joseph~A. Konstan.
\newblock Interactive content diversity and user exploration in online movie recommenders: A field experiment.
\newblock \emph{International Journal of Human–Computer Interaction}, 40\penalty0 (22):\penalty0 7233--7247, 2024.
\newblock URL \url{https://doi.org/10.1080/10447318.2023.2262796}.

\bibitem[Wang et~al.({\natexlab{a}})Wang, Shivanna, Cheng, Jain, Lin, Hong, and Chi]{DCNV2}
Ruoxi Wang, Rakesh Shivanna, Derek Cheng, Sagar Jain, Dong Lin, Lichan Hong, and Ed~Chi.
\newblock Dcn v2: Improved deep \& cross network and practical lessons for web-scale learning to rank systems.
\newblock In \emph{Proceedings of the Web Conference 2021}, page 1785–1797, New York, NY, USA, {\natexlab{a}}. Association for Computing Machinery.
\newblock URL \url{https://doi.org/10.1145/3442381.3450078}.

\bibitem[Wang et~al.({\natexlab{b}})Wang, He, Wang, Feng, and Chua]{ngcf}
Xiang Wang, Xiangnan He, Meng Wang, Fuli Feng, and Tat-Seng Chua.
\newblock Neural graph collaborative filtering.
\newblock page 165–174. ACM, {\natexlab{b}}.
\newblock \doi{10.1145/3331184.3331267}.

\bibitem[Wang et~al.(2024)Wang, Sun, Yuan, and Chen]{wang2024field}
Yitong Wang, Tianshu Sun, Zhe Yuan, and AJ~Yuan Chen.
\newblock How recommendation affects customer search: A field experiment.
\newblock \emph{Information Systems Research}, 36\penalty0 (1):\penalty0 85--106, 2024.

\bibitem[Wu et~al.()Wu, Wang, Feng, He, Chen, Lian, and Xie]{sgl}
Jiancan Wu, Xiang Wang, Fuli Feng, Xiangnan He, Liang Chen, Jianxun Lian, and Xing Xie.
\newblock Self-supervised graph learning for recommendation.
\newblock In \emph{Proceedings of the 44th International ACM SIGIR Conference on Research and Development in Information Retrieval}, page 726–735, New York, NY, USA. Association for Computing Machinery.
\newblock URL \url{https://doi.org/10.1145/3404835.3462862}.

\bibitem[Yao et~al.(2021)Yao, Halpern, Thain, Wang, Lee, Prost, Chi, Chen, and Beutel]{FERRERO_EFFECTS}
Sirui Yao, Yoni Halpern, Nithum Thain, Xuezhi Wang, Kang Lee, Flavien Prost, Ed~H. Chi, Jilin Chen, and Alex Beutel.
\newblock Measuring recommender system effects with simulated users, 2021.
\newblock URL \url{https://arxiv.org/abs/2101.04526}.

\bibitem[Yi et~al.(2022)Yi, Kim, and Ju]{yi2022recommendation}
Sangyoon Yi, Dongyeon Kim, and Jaehyeon Ju.
\newblock Recommendation technologies and consumption diversity: An experimental study on product recommendations, consumer search, and sales diversity.
\newblock \emph{Technological Forecasting and Social Change}, 178:\penalty0 121486, 2022.

\bibitem[Zeng et~al.(2020)Zeng, Fang, and Si]{zeng2020behavioral}
Yan Zeng, Yuan Fang, and Luo Si.
\newblock User behavior modeling in recommender system simulations.
\newblock In \emph{Proceedings of the 43rd International ACM SIGIR Conference on Research and Development in Information Retrieval (SIGIR ’20)}, pages 1509--1512, 2020.

\bibitem[Zeng et~al.(2021)Zeng, Chen, Sun, Fang, and Si]{zeng2021recwalk}
Yan Zeng, Zhenzhong Chen, Jianyuan Sun, Yuan Fang, and Luo Si.
\newblock Recwalk: Nearly unbiased walk-based recommendation via markov chains.
\newblock In \emph{Proceedings of the 14th ACM International Conference on Web Search and Data Mining (WSDM ’21)}, pages 367--375, 2021.
\newblock URL \url{https://dl.acm.org/doi/10.1145/3437963.3441818}.

\bibitem[Zhao et~al.(2022)Zhao, Hou, Pan, Yang, Zhang, Lin, Zhang, Bian, Tang, Sun, Chen, Xu, Zhang, Tian, Tian, Mu, Fan, Chen, and Wen]{recbole2.0}
Wayne~Xin Zhao, Yupeng Hou, Xingyu Pan, Chen Yang, Zeyu Zhang, Zihan Lin, Jingsen Zhang, Shuqing Bian, Jiakai Tang, Wenqi Sun, Yushuo Chen, Lanling Xu, Gaowei Zhang, Zhen Tian, Changxin Tian, Shanlei Mu, Xinyan Fan, Xu~Chen, and Ji{-}Rong Wen.
\newblock Recbole 2.0: Towards a more up-to-date recommendation library.
\newblock \emph{arXiv preprint arXiv:2206.07351}, 2022.

\bibitem[Zheng et~al.()Zheng, Lu, Jiang, Zhang, and Yu]{spectralCF}
Lei Zheng, Chun-Ta Lu, Fei Jiang, Jiawei Zhang, and Philip~S. Yu.
\newblock Spectral collaborative filtering.
\newblock In \emph{Proceedings of the 12th ACM Conference on Recommender Systems}, page 311–319, New York, NY, USA. Association for Computing Machinery.
\newblock URL \url{https://doi.org/10.1145/3240323.3240343}.

\bibitem[Zhou et~al.()Zhou, Kuscsik, Liu, Medo, Wakeling, and Zhang]{RARITY_1}
Tao Zhou, Zoltán Kuscsik, Jian-Guo Liu, Matúš Medo, Joseph~Rushton Wakeling, and Yi-Cheng Zhang.
\newblock Solving the apparent diversity-accuracy dilemma of recommender systems.
\newblock \emph{Proceedings of the National Academy of Sciences}, 107\penalty0 (10):\penalty0 4511--4515.
\newblock URL \url{https://www.pnas.org/doi/abs/10.1073/pnas.1000488107}.

\end{thebibliography}

\appendix
\section{Supplementary Results}
\label{sec:appA}
\begin{figure}[ht!] 
    \label{alg:simulation}
    \centering
    \includegraphics[width=\linewidth]{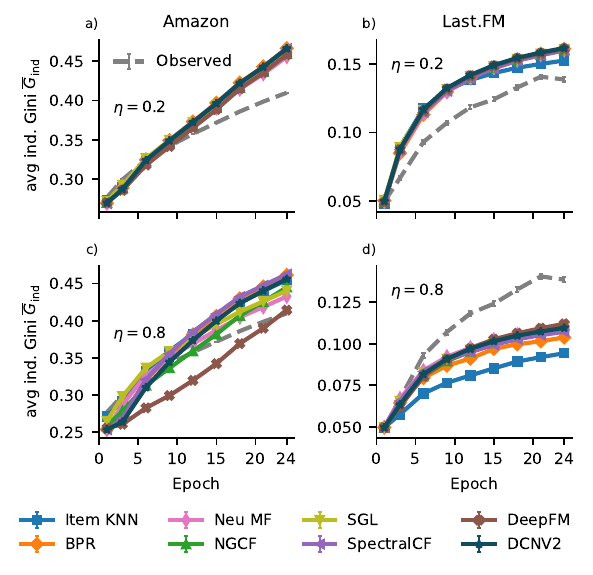}
    \caption{Evolution of the average individual Gini at different adoption rates. (a-b) $\eta=0.2$. (c-d) $\eta=0.8$.}
    \label{fig1:ind_gini_epochs}
\end{figure}

Figure \ref{fig1:ind_gini_epochs} show the average individual Gini coefficient across epochs, for all considered models. For each dataset, results are reported for adoption rates $\eta=0.2$ and $\eta=0.8$. 

Figure \ref{fig1:jaccard_temp_appendix} shows an analogous visualization based on the Jaccard similarity.
Results are shown for adoption rates $\eta=0.2$ and $\eta=0.8$.

\begin{figure}[ht!] 
    \label{alg:simulation}
    \centering
    \includegraphics[width=\linewidth]{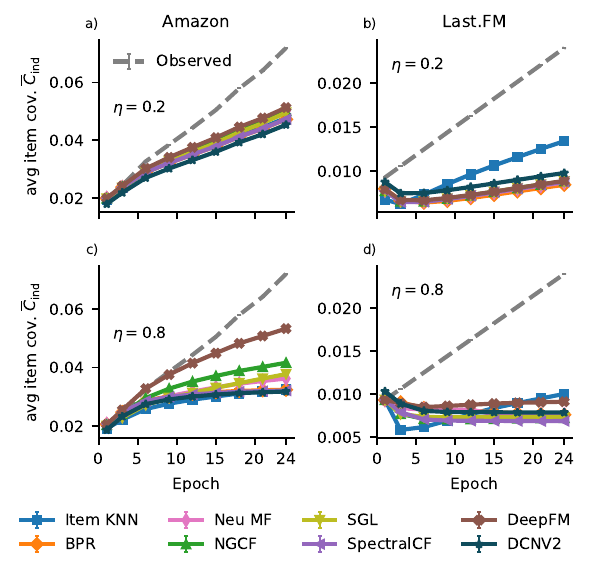}
    \caption{Evolution of the average individual item coverage at different adoption rates. (a-b) $\eta=0.2$. (c-d) $\eta=0.8$.}
    \label{fig1:user_cov}
\end{figure}

\begin{figure}[ht!]
    \centering
    \includegraphics[width=\linewidth]{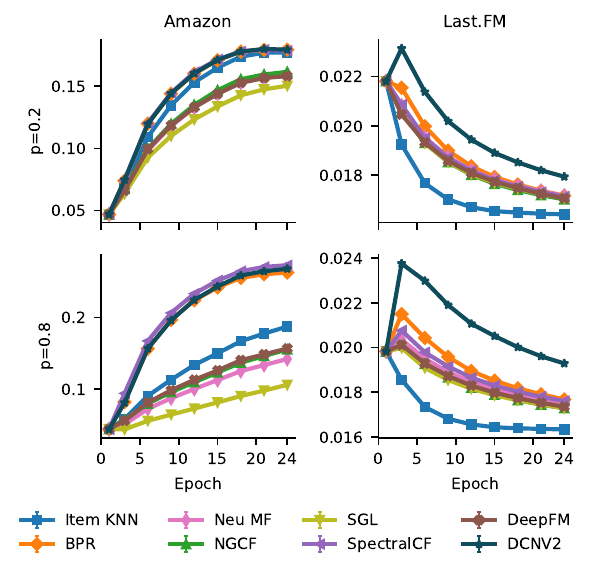}
    \caption{Evolution of the Jaccard similarity at different adoption rates.}
    \label{fig1:jaccard_temp_appendix}
\end{figure}

\newpage
\section{Recommender Systems}
\label{sec:appB}

Table \ref{tab:rec-results-datasets} reports the accuracy of the models evaluated through the temporal holdout process. 
Results are reported separately for each dataset, and the model achieving the best performance during the initialization phase is identified.

During the initial testing phase, a hyperparameter optimization step is also performed for each dataset. Tables \ref{tab:best-hparams_amazon} and \ref{tab:best-hparams_lastfm} report the best hyperparameter configurations identified.

\begin{table*}[htb!]
  \centering
  \small
  \caption{Performance comparison of recommendation systems across datasets.
  Best results within each dataset are in \textbf{bold}.}
  \label{tab:rec-results-datasets}
  \begin{threeparttable}
    \sisetup{detect-weight=true, detect-family=true}
    \resizebox{\textwidth}{!}{%
    \begin{tabular}{l
                    S[table-format=1.4] S[table-format=1.4] S[table-format=1.4] S[table-format=1.4]
                    S[table-format=1.4] S[table-format=1.4] S[table-format=1.4] S[table-format=1.4]}
      \toprule
      & \multicolumn{4}{c}{Amazon} & \multicolumn{4}{c}{Last.fm} \\
      \cmidrule(lr){2-5} \cmidrule(lr){6-9}
      Model 
      & {N@10} & {P@10} & {R@10} & {ItemCov}
      & {N@10} & {P@10} & {R@10} & {ItemCov} \\
      \midrule
      SGL 
      & \textbf{0.3597} & \textbf{0.1514} & \textbf{0.2794} & \textbf{0.7053}
      & 0.3786 & 0.712 & 0.1292 & 0.1865 \\

      Neu MF
      & 0.2987 & 0.1502 & 0.2891 & 0.4051
      & 0.7391 & 0.7233 & 0.1262 & 0.1529 \\

      BPR
      & 0.2496 & 0.1101 & 0.2365 & 0.0745
      & \textbf{0.7582} & \textbf{0.7327} & \textbf{0.1225} & \textbf{0.1405} \\

      NGCF
      & 0.2303 & 0.1121 & 0.2332 & 0.3899
      & 0.666 & 0.6394 & 0.0976 & 0.164 \\

      SPECTRAL CF
      & 0.1996 & 0.0951 & 0.2029 & 0.0173
      & 0.1659 & 0.0873 & 0.1934 & 0.5721 \\

      DCN V2
      & 0.1736 & 0.0857 & 0.1827 & 0.0414
      & 0.4504 & 0.4426 & 0.064 & 0.0727 \\

      Item KNN
      & 0.1199 & 0.0769 & 0.1147 & 0.6059
      & 0.5901 & 0.5526 & 0.0768 & 0.2563 \\

      DeepFM
      & 0.1083 & 0.0561 & 0.1209 & 0.5611
      & 0.7015 & 0.6948 & 0.135 & 0.1227 \\

      \bottomrule
    \end{tabular}%
    }
    \begin{tablenotes}
      \footnotesize
      \item N@10 = NDCG@10, P@10 = Precision@10, R@10 = Recall@10. All metrics are averaged over users; higher values indicate better performance.
    \end{tablenotes}
  \end{threeparttable}
\end{table*}

\begin{table}[t]
\centering
\small
\setlength{\tabcolsep}{3pt}
\caption{Best hyperparameters used in our experiments for the Amazon dataset}
\label{tab:best-hparams_amazon}
\begin{tabular}{l|c|p{0.8\textwidth}}
\hline
\textbf{Model} & \textbf{Training} & \textbf{Model-specific hyperparameters} \\
\hline

CF-KNN (item) &
$E{=}1$ &
$k{=}100$, shrink$=0.0$ \\

BPR &
$E{=}50$, lr$=0.001$ &
-- \\

NeuMF &
$E{=}50$, lr$=0.004$ &
dropout$=0.2$, MLP$=[64,32,16]$ \\

NGCF &
$E{=}50$, lr$=0.02$ &
dim$=32$, layers$=[64,64,64]$, node-drop$=0.2$, msg-drop$=0.01$, $\lambda{=}0.002$ \\

SGL &
$E{=}50$, lr$=0.03$ &
dim$=128$, $L{=}3$, $\lambda{=}0.002$, $\tau{=}0.12$, $\alpha{=}0.03$, drop-ratio$=0.2$, aug=ND \\

SpectralCF &
$E{=}50$, lr$=0.0045$ &
$L{=}3$, $\lambda{=}0.0005$ \\

DeepFM &
$E{=}50$, lr$=0.02$ &
dim$=32$, MLP$=[64,64]$, dropout$=0.4$ \\

DCN-v2 &
$E{=}50$, lr$=0.008$ &
dim$=16$, MLP$=[768,768]$, dropout$=0.1$, $\lambda{=}0.0005$, cross-layers$=3$, structure=stacked, experts$=3$, low-rank$=128$ \\

\hline
\end{tabular}
\end{table}

\begin{table}[t]
\centering
\small
\setlength{\tabcolsep}{3pt}
\caption{Best hyperparameters used in our experiments for the Last.fm dataset}
\label{tab:best-hparams_lastfm}
\begin{tabular}{l|c|p{0.8\textwidth}}
\hline
\textbf{Model} & \textbf{Training} & \textbf{Model-specific hyperparameters} \\
\hline

CF-KNN (item) &
$E{=}1$ &
$k{=}150$, shrink$=0.0$ \\

BPR &
$E{=}50$, lr$=0.0001$ &
-- \\

NeuMF &
$E{=}50$, lr$=0.004$ &
dropout$=0.1$, MLP$=[64,64,64]$ \\

NGCF &
$E{=}50$, lr$=0.0002$ &
dim$=32$, layers$=[64,64,64]$, node-drop$=0.2$, msg-drop$=0.01$, $\lambda{=}0.002$ \\

SGL &
$E{=}50$, lr$=0.003$ &
dim$=64$, $L{=}3$, $\lambda{=}0.002$, $\tau{=}0.12$, $\alpha{=}0.03$, drop-ratio$=0.2$, aug=ND \\

SpectralCF &
$E{=}50$, lr$=0.006$ &
$L{=}3$, $\lambda{=}0.0005$ \\

DeepFM &
$E{=}50$, lr$=0.003$ &
dim$=32$, MLP$=[64,64]$, dropout$=0.4$ \\

DCN-v2 &
$E{=}50$, lr$=0.0008$ &
dim$=16$, MLP$=[768,768]$, dropout$=0.1$, $\lambda{=}0.0005$, cross-layers$=3$, structure=stacked, experts$=3$, low-rank$=128$ \\

\hline
\end{tabular}
\end{table}

\section{Simulation Algorithms}
Algorithm 1 defines the overall framework and drives the simulation by integrating all the necessary components and establishing the main loop, which outputs the simulation epochs. Algorithm 2 provides a detailed description of how a user, when prompted to interact, simulates the selection of an item from the recommendation basket, either by relying on the user’s personal model or by sampling from the recommendations generated by the model.

\begin{algorithm}[ht]
\caption{Simulation framework}
\label{alg:simulation}
\begin{algorithmic}[1]
\REQUIRE Historical interaction data $D$; total epochs $E$; last initialization epoch $E_{\text{last-init}}$; retraining interval $\Delta$; cold-start cut $t_0$
\ENSURE Simulated interaction dataset $D_{\text{post}}$

\STATE $D_{\text{init}} \gets \textsc{ColdStart}(D, t_0)$ \COMMENT{Initialization prefix used for calibration and first training}
\STATE $R \gets \textsc{AlgorithmTraining}(D_{\text{init}})$ \COMMENT{Recommender scoring model}
\STATE $\hat{R} \gets \textsc{ChoiceModelSetUp}(D_{\text{init}})$ \COMMENT{Autonomous user choice model}
\STATE $D_{\text{post}} \gets D_{\text{init}}$
\STATE $e \gets E_{\text{last-init}} + 1$
\STATE $E_{\text{last-train}} \gets E_{\text{last-init}}$

\WHILE{$e \leq E$}
    \STATE $S_e \gets \textsc{GetStepsInEpoch}(e)$
    \FORALL{$t \in S_e$}
        \STATE $U_t \gets \textsc{GetAwakenUsers}(e,t)$
        \FORALL{$u \in U_t$}
            \STATE $b_u \gets \textsc{GetBasketSize}(u,t)$
            \STATE $I_u \gets \textsc{ItemSelection}(u, b_u, R, \hat{R}, e, t)$
            \FORALL{$i \in I_u$}
                \STATE $D_{\text{post}} \gets D_{\text{post}} \cup \{(u,i,t)\}$
            \ENDFOR
        \ENDFOR
    \ENDFOR

    \IF{$e - E_{\text{last-train}} \geq \Delta$}
        \STATE $R \gets \textsc{AlgorithmTraining}(D_{\text{post}})$
        \STATE $\hat{R} \gets \textsc{ChoiceModelSetUp}(D_{\text{post}})$
        \STATE $E_{\text{last-train}} \gets e$
    \ENDIF

    \STATE $e \gets e+1$
\ENDWHILE
\end{algorithmic}
\end{algorithm}

\begin{algorithm}[t]
\caption{\textsc{ItemSelection} for user $u$}\label{alg:itemsel}
\begin{algorithmic}[1]
\REQUIRE User $u$; basket size $b_u$; recommender model $R$; choice model $\hat{R}$; epoch $e$; step $t$
\STATE \textbf{Hyperparameters:} adoption rate $\eta$; recommendation list size $K$
\ENSURE Basket of selected items $I_u$

\STATE $C_{u,e} \gets \textsc{GetCandidateSet}(u,e)$
\STATE $K_{u,t} \gets \textsc{TopK}(R,u,t,K)$
\STATE $I_u \gets \emptyset$

\FOR{$j=1$ \TO $b_u$}
    \IF{$\textsc{Bernoulli}(\eta)=1$}
        \STATE $i \gets \textsc{SampleFromRecList}(K_{u,t}, R(u,\cdot,t))$ \COMMENT{Score-proportional sampling on $K_{u,t}$}
    \ELSE
        \STATE $i \gets \textsc{SampleFromCandidateSet}(C_{u,e}, \hat{R}(u,\cdot,t))$ \COMMENT{Softmax choice on $C_{u,e}$}
    \ENDIF
    \STATE $I_u \gets I_u \cup \{i\}$
\ENDFOR

\RETURN $I_u$
\end{algorithmic}
\end{algorithm}

\end{document}